\documentstyle[sprocl,epsfig]{article}

\def\beq{\begin{equation}}
\def\eeq#1{\label{#1}\end{equation}}
\def\eeqn{\end{equation}}

\def\beqa{\begin{eqnarray}}
\def\eeqa#1{\label{#1}\end{eqnarray}}
\def\eeqan{\end{eqnarray}}

\let\bar=\overbar

\def\Dslash{\not{\hbox{\kern-4pt $D$}}}
\def\dslash{\not{\hbox{\kern-2pt $\del$}}}

\def\ee{e^+e^-}

\def\msb{{\bar{\ssstyle M \kern -1pt S}}}

\def\Cpp{C\verb-++-}

\def\Title#1{\begin{center} {\Large #1 } \end{center}}
\def\Author#1{\begin{center}{ \sc #1} \end{center}}
\def\Address#1{\begin{center}{ \it #1} \end{center}}

\def\submit#1{\begin{center} #1 \end{center}}
\def\doeack{\footnote{Work supported by the Department of Energy,
                     contract DE--AC03--76SF00515.}}
\def\SLAC{Stanford Linear Accelerator Center\\
    Stanford University, Stanford, California 94309 USA}
\newcommand\pubblock{\rightline{\begin{tabular}{l} 
         SLAC-PUB-8290\\ October 1999 \end{tabular}}}
\newenvironment{Abstract}{\begin{quotation} \begin{center}
                       ABSTRACT
     \end{center}\bigskip  }{\end{quotation}}

\begin{document}
\begin{titlepage}
\pubblock

\vfill
\Title{Pandora: an Object-Oriented 
Event Generator for Linear Collider Physics}
\vfill
\Author{Michael E. Peskin\doeack}
\Address{\SLAC}
\vfill
\begin{Abstract}
I desribe a new event generator, pandora, which uses the \Cpp\ class 
structure
to allow a modular treatment of beams and particle production and decay.
\end{Abstract}
\vfill
\submit{presented at the International Workshop on Linear Colliders\\
          Sitges, Barcelona, Spain, 28 April -- 5 May 1999}

\vfill
\end{titlepage}
\def\thefootnote{\fnsymbol{footnote}}
\setcounter{footnote}{0}

\hbox to \hsize{\null}
\newpage
\setcounter{page}{1}

\title{PANDORA: AN OBJECT-ORIENTED EVENT GENERATOR FOR LINEAR COLLIDER PHYSICS}

\author{MICHAEL E. PESKIN}

\address{Stanford Linear Accelerator Center\\
    Stanford University, Stanford, California 94309 USA}

\maketitle\abstracts{I describe an new event generator, pandora,
      which uses the \Cpp\ class structure to allow a 
       modular treatment of beams and particle  production and decay.}

\section{Introduction}

The ideal LC event generator  needs to fulfill a number of requirements.
It should provide the basic standard and nonstandard processes 
in $\ee$ annihilation.  It should also include beamstrahlung and initial
state radiation, include initial-state polarization, and include the
full set of final-state correlations associated with polarization effects.
It should also allow input of any parton-level cross section
 and should be able to generate events from this cross section with reasonable
efficiency.

  As a step toward these goals, I would like to introduce a new
generator, called pandora.
  The idea of pandora is to package the various stages
of a LC event as distinct \Cpp\ classes which interact through a simple 
interface. This makes it possible to include new parton-level processes
in a simple way, allowing the larger system to take care of the initial
beams and final interactions.

Pandora is implemented in \Cpp\ as a \verb+pandora+ class, whose constructor
depends on two \verb+beam+ classes and one \verb+process+ class.  Utility
classes provide the distributions for $W$, $Z$, and top quark decay.
All of the classes make use of an \verb+LVector+ (Lorentz vector) class
which provides a variety of 4-vector operations.  A complete main program 
for parton-level $\ee\to t \bar t$ events is shown in 
Table~\ref{tab:program}.  The current software distribution and 
documentation can be found at ref. 1.

\begin{table}[t]
\begin{center}
\caption{A simple program with pandora}
\label{tab:program}
\begin{verbatim}
#include "pandora.h"
#include "eetottbar.h"
#include "beams.h"
int main(){
    double ECM = 500.0
    ebeam B1(ECM/2.0, electron, electron, 0.9, 1, 1, NLC500);
    ebeam B2(ECM/2.0, positron, positron, 0.0, 1, 1, NLC500);
    eetottbar R;
    pandora P(B1,B2,R);
    int Ncalls = 100000;
    VegasGrid V = P.prepare(Ncalls);
    for (int i =1;i <= 10;i++)
       {LEvent LE = P.getEvent(V);cout << LE ;}
    return 0;}
\end{verbatim}
\end{center}
\end{table}

\section{Event selection}

The basic concept of pandora is to represent a cross section for a complete
$\ee$ process---from the initial electron and positron to the final 
partons---as an integral over $N$ variables $x_i$ which run over $[0,1]$.
The $x_i$ parametrize all relevant variables in the process, from 
beamstrahlung energy loss to resonance decay angles.  The expression
$d^n \sigma/dx^n$ is handed to a general purpose program to select weight-1
events.  Then each chosen value of $\{ x_i\}$ is converted to an event
handed back in a standard format specified by  the \verb+LEvent+ class.

In pandora,
event selection is done using the VEGAS algorithm to optimize a grid in the
$N$-dimensional space and then choosing weight-1 events in the metric 
defined by this grid.  This is the algorithm used in Kawabata's BASES/SPRING
program.\cite{Kawabata}  VEGAS does do useful adaptation to the function
being integrated, but the algorithm is inefficient if the peaks of the 
function being integrated are not aligned with the grid. Methods
recently proposed by Ohl~\cite{Ohl} and Jadach~\cite{Jadach} may ameliorate
this problem.  For the moment, we accept a loss in speed of event generation
as the price of generality.

\section{Beam class}

The parametrizations of beamstrahlung
 and initial-state
radiation are contained in classes which are derived from the abstract
class \verb+beam+. The formulae for beamstrahlung 
used in pandora are based on the `consistent
Yokoya-Chen' approximation explained in ref. 5.  They are somewhat
simpler than earlier analytic formulae in the literature and agree just as 
well (or poorly) with simulation data.   Both $\ee$ and $e^-e^-$ (the latter
for round beams only) are considered.  The \verb+ebeam+ 
class constructor used in 
Table~\ref{tab:program} takes as arguments, the nominal beam energy, the
beam species, the species initiating the considered
reaction (which might, for example,
 be a photon
from an electron beam), the beam polarization, flags for the inclusion of 
beamstrahlung and initial-state radiation, and the name of a reference 
machine design.  A more complicated constructor allows input of arbitrary
beam parameters.

\section{Process class}

The reaction cross sections and decay distributions are contained
in classes which are derived from the abstract class \verb+process+.
A subclass of \verb+process+ must define three functions, which give
the allowed domain of the variables $x_i$, the value of the differential
cross section, and the event corresponding to a chosen set  $\{ x_i\}$.
The definition of \verb+process+ is given in Table~\ref{tab:process}.
The differential cross section is returned as a 4-component 
vector for the four possible
orientations of initial helicities.  

\begin{table}[t]
\begin{center}
\caption{Definition of the pandora process class}
\label{tab:process}
\begin{verbatim}
class process{ public:
  process(int N): n(N){}
  virtual int validEvent(DVector & X, double s, double beta) = 0;
  virtual DVector crosssection(DVector & X, double  s) = 0; 
  virtual LEvent buildEvent(DVector & X, double s) = 0;
               private:
  int n;  /* number of integration variables */   }; 
\end{verbatim}
\end{center}
\end{table}

\begin{figure}
\begin{center}
\leavevmode
{\epsfxsize=2.00truein \epsfbox{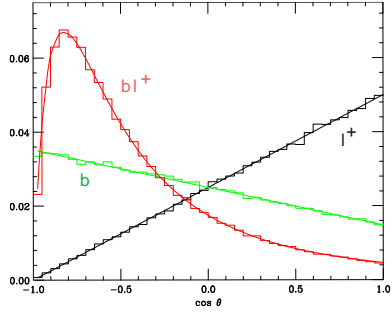}}
\end{center}
 \caption{Distribution of decay angles of
 the $\ell^+$ and $b$ in top decay, measured 
with respect to the top spin direction, and the angle between the $\ell^+$
and $b$, comparing pandora results to lowest-order theory.}
\label{fig:tangles}
\end{figure}

More generally, pandora returns amplitudes as
matrices indexed by polarization, which
are multiplied to obtain cross sections with full spin correlations. As an 
example, Figure~\ref{fig:tangles} shows the distributions of 
top quark decay angles 
returned by pandora. The helicity  amplitudes for particle production and
decay
are the basic raw materials 
for linear collider physics studies. It is part of my plan to provide a 
compilation of the amplitudes needed for the most imporant LC 
processes.

\section{Interface to PYTHIA}

Pandora returns parton-level final states and does not carry out hadronization.
This is acceptable if the final states can be hadronized by a general-purpose
simulation program such as PYTHIA.\cite{Pythia}
  Since PYTHIA lives in the FORTRAN world,
its coupling to pandora must be somewhat inelegant.  However, Barklow and 
Iwasaki have written a general interface which includes pandora processes as 
subprocesses in PYTHIA event generation.  Tau leptons are decayed using 
TAUOLA,\cite{Tauola} taking account of their longitudinal polarization,
before the event is hadronized.  To facilitate the interface to PYTHIA, pandora
returns in the \verb+LEvent+ class the color contractions of the final 
partons and the order in which partons are to be taken in pairs to compute
QCD showers.  This interface, \verb+PANDORA_PYTHIA+, is avaiable from the 
pandora Web site.\cite{pandorasite}

\section*{Acknowledgments}
I am grateful
 to Tim Barklow and Masako Iwasaki for their many contributions to this 
project, and to David Gerdes and Ester Ruiz Morales for helpful
criticism.    This 
work was supported by the US Department of Energy under contract
DE--AC03--76SF00515.

\end{document}